\documentclass[a4paper]{jpconf}
\usepackage{graphicx}
\usepackage{subfigure}
\usepackage{hyperref}
\pdfoutput=1

\newcommand{\ttH}{\ensuremath{t\bar{t}H}}
\newcommand{\bbbar}{\ensuremath{b\bar{b}}}
\newcommand{\ttbar}{\ensuremath{t\bar{t}}}

\begin{document}

\title{Search for $t\bar{t}H$~production at the LHC}

\author{Mark Owen \\ On behalf of the ATLAS \& CMS Collaborations}

\address{The University of Glasgow, Glasgow,  G12 8QQ, UK}

\ead{markowen@cern.ch}

\begin{abstract}
The searches for the production of the Higgs boson associated with a pair of top
quarks in the ATLAS and CMS experiments are presented. 
The searches use a range of final states sensitive to the Higgs boson decaying
into $b$-quark pairs, pairs of vector bosons, pairs of taus and pairs of photons.
All the searches use $pp$~collision data at
$\sqrt{s} = 8$~TeV collected with the ATLAS and CMS detectors at the LHC in 2012
and some analyses also include the data collected at $\sqrt{s} = 7$~TeV in 2011.
The searches in the $\bbbar$~and $\gamma\gamma$~channels observe no
excess of events relative to the background expectation, while the CMS search
using multi-lepton events observes an excess of slightly more than $3$~standard
deviations over the background expectation.
\end{abstract}

\section{Introduction}

The discovery of a Higgs boson was reported by the ATLAS~\cite{atlasdetectorpaper} and CMS~\cite{Chatrchyan:2008zzk} collaborations
in July 2012~\cite{2012gk,2012gu}.
Determining whether the new boson behaves as predicted by the Standard Model (SM) is of primary importance
and in particular it is vital to observe it in as many production and decay modes as possible.

The production of the Higgs boson in association with a pair of top quarks ($\ttH$)~provides direct sensitivity to
the large Yukawa coupling between the top quark and the Higgs boson~\cite{Beenakker:2001rj}. This is in contrast to the gluon fusion production mode,
where the sensitivity to the top quark coupling is only via loop effects.

The decay modes of the Higgs boson, $H\to b\bar{b}$, $H\to W^+W^-$, $H\to ZZ$, $H \to \tau^+\tau^-$~and
$H\to \gamma\gamma$~give a range of different final states in which to search for $\ttH$~production. The searches
from the ATLAS and CMS collaborations using these different signatures are discussed below. The searches
use the full dataset collected during $pp$~collisions at the LHC in 2012 at $\sqrt{s}=8$~TeV
and in some cases also include the data collected in 2011 at $\sqrt{s}=7$~TeV.

\section{$\ttH$~search with $H\to b\bar{b}$}
\label{sec-bb}

For a Higgs boson with mass $m_H=125$~GeV the SM predicts that the dominant decay mode is into a pair of $b$-quarks.
This decay mode therefore provides the largest number of signal events for the $\ttH$~process. However, the large background
from the production of $\ttbar$~in association with additional jets makes this a challenging signature to separate from the background.
Both ATLAS and CMS have searched for $\ttH$~production using the $\bbbar$~decay mode~\cite{ATLASConf2014-011,Khachatryan:2014qaa,CMSPasHig14-010}.

The dominant background to the signal process comes from the production of $\ttbar$~in association with additional jets and in the high purity signal regions
the additional jets are dominated by jets originating from heavy flavour quarks ($c$~or $b$). Both collaborations estimate this background using MC simulation,
reweighted to match measurements of the $\ttbar$~differential cross-section. Additional small contributions to the background come from the production
of vector bosons plus jets, single top-quark production and QCD multi-jet production.

\begin{figure}[b]
\begin{center}
\subfigure[]{
\includegraphics[width=0.34\textwidth]{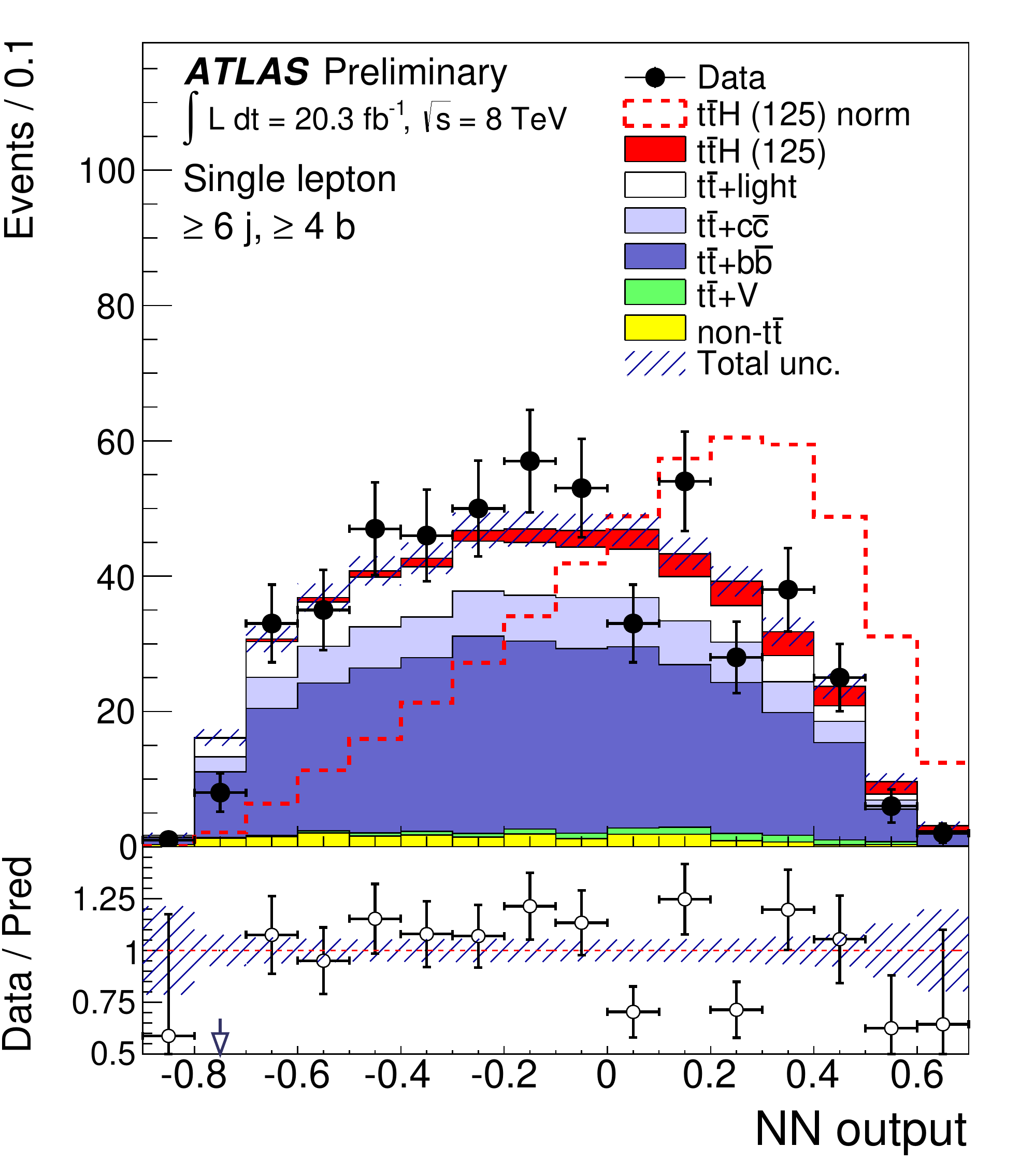}}
\subfigure[]{
\includegraphics[width=0.36\textwidth]{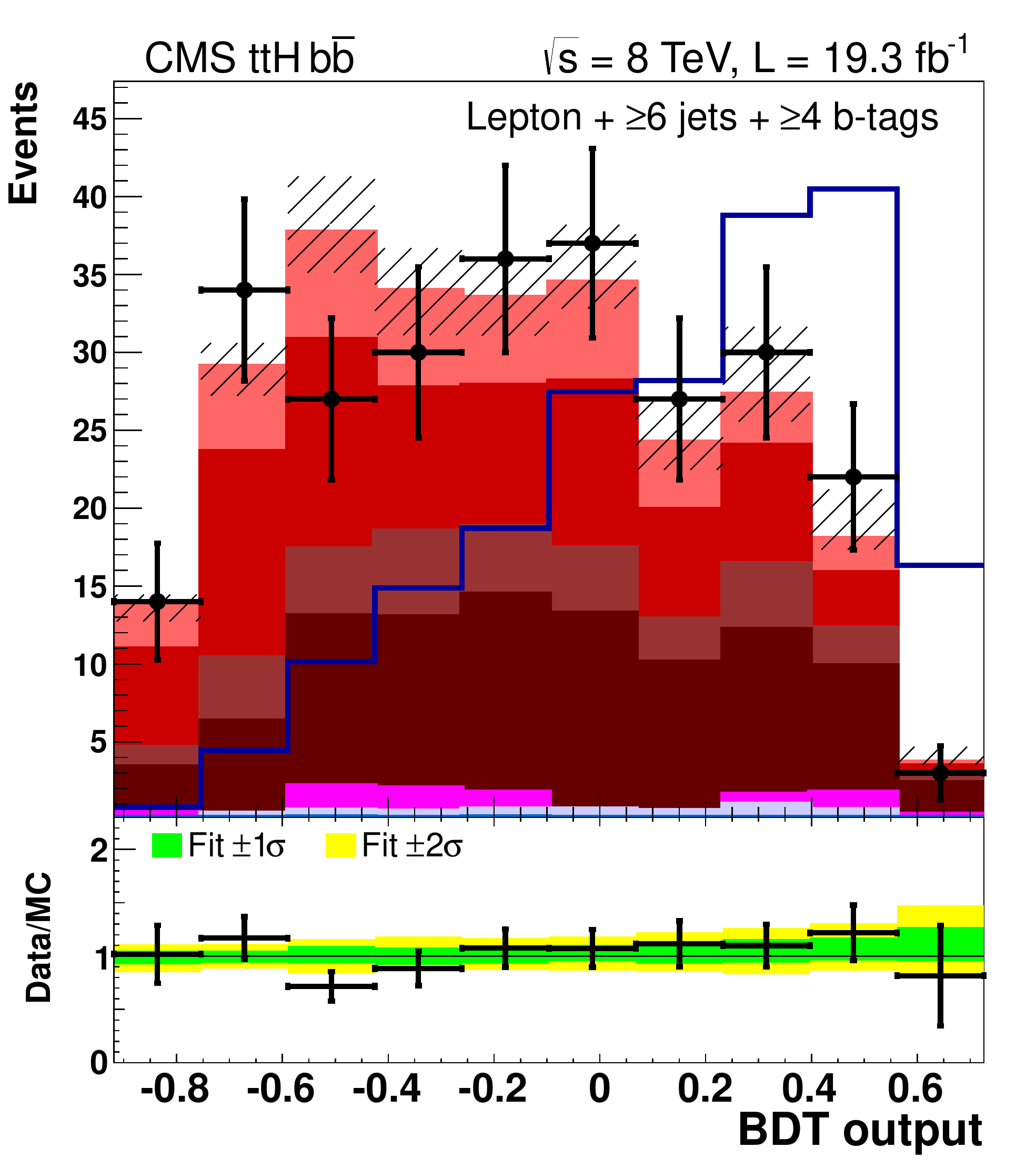}
\includegraphics[width=0.18\textwidth]{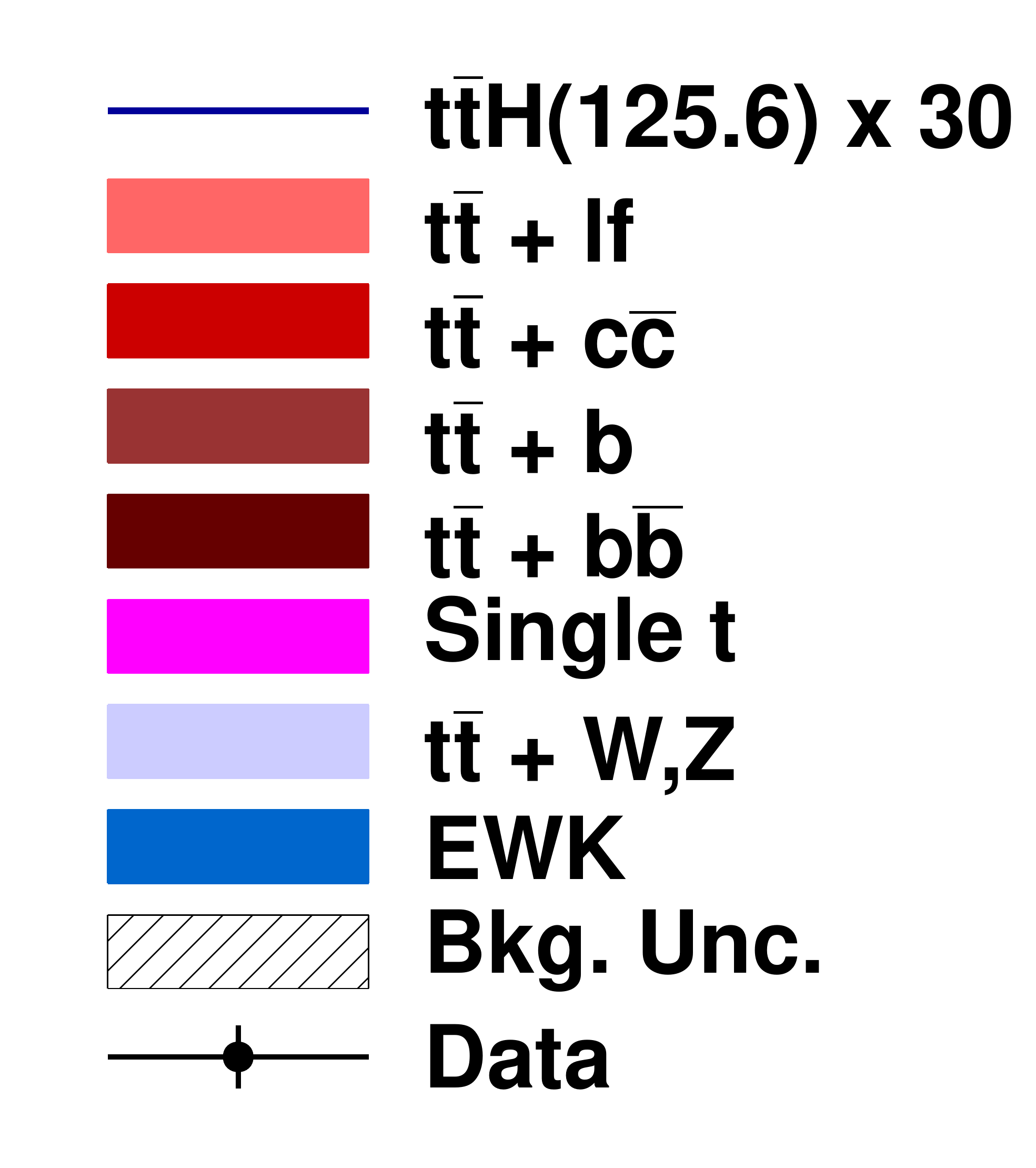}
}
\end{center}
\caption{\label{fig-bb-mvoutput}Comparison of the data to the sum of the expected backgrounds for events with at least 6 jets, at least 4 of which are identified as $b$-jets, for (a)
the NN output for the ATLAS $H\to\bbbar$~analysis and (b) the BDT output for the CMS $H\to\bbbar$~analysis.}
\end{figure}

The top quark decays in the SM to a $W$~boson and a $b$-quark almost $100\%$~of the time. 
The analyses use the decay modes of the top-quark pairs where at least one of the top quarks has decayed into a lepton and
neutrino (via a $W$~boson). This gives two final states, the lepton plus jets final state, where one $W$~boson has decayed leptonically
and the other has decayed into a pair of quarks, and the dilepton final state, where both $W$~bosons decay leptonically.
Both collaborations split the selected samples according to the number of observed jets and the number of those jets that have
been identified as originating from $b$-quarks ($b$-jets). The $b$-jet identification algorithms have typical efficiencies to correctly identify $b$-jets of $70\%$~for both experiments, with the ATLAS (CMS) algorithm rejecting $99\%\ (98\%)$~of light quark / gluon initiated jets.
In the highest-purity signal regions (with at least 4 $b$-jets), the signal-to-background ratio is still low, for example $3.8\%$~for the ATLAS lepton plus jets selection.
Therefore, both collaborations employ multivariate techniques to separate the signal from the background. ATLAS uses a Neural Network (NN) trained using kinematic variables,
while CMS uses Boosted Decision Trees (BDT), where in addition to kinematic variables, information from the $b$-jet identification algorithm is also used.
The data
are in good agreement with the background expectation, as illustrated in Figure~\ref{fig-bb-mvoutput} where the NN and BDT discriminants
are shown for events with at least 6 jets, 4 of which are identified as $b$-jets.

Each collaboration performs a combined fit to the discriminants in the different jet / $b$-jet multiplicity regions. The dominant systematic uncertainties originate from the modelling
of the $\ttbar +$jets background and the understanding of the efficiency and rejection power of the $b$-jet identification algorithms. The resulting cross-section measurement, $\sigma$, is expressed relative to the SM
cross-section, $\mu = \sigma / \sigma_{\mathrm{SM}}$; ATLAS measuring $\mu=1.7\pm1.4$~and CMS measuring $\mu=0.7\pm1.9$. Neither experiment observes a significant deviation
relative to the background only hypothesis and both experiments also report $95\%$~confidence level (CL) upper limits on $\mu$. The observed (expected) limit from ATLAS
is 4.6 (2.6) times the SM cross-section, while CMS finds observed (expected) limits of 4.1 (3.5) times the SM cross-section.

The CMS collaboration has also used a matrix element method to perform the search for $\ttH$~with the Higgs boson decaying into $b$-quark pairs~\cite{CMSPasHig14-010}.
The matrix element method uses leading-order matrix elements from the $\ttH$~signal process and $\ttbar + \bbbar$~background process to provide separation between signal
and background. The information from the $b$-tagging algorithms is added to the output of the matrix element method in order to improve the separation against $\ttbar$~production with
additional jets originating from light flavour quarks / gluons. This analysis reaches an observed (expected) limit of 3.3 (2.9) times the SM cross-section. The improvement in the expected limit
relative to the BDT-based result demonstrates the improvements that are possible for the discrimination between the signal and backgrounds.

\section{$\ttH$~search with $H\to \gamma\gamma$}
\label{sec-gg}

In the SM, the branching ratio for the decay $H \to \gamma\gamma$~is $2.28 \times 10^{-3}$~for $m_H = 125$~GeV. Despite the small rate, the diphoton final state
provides a clean signal with very good resolution of $m_{\gamma\gamma}$, making this channel promising for the $\ttH$~search. The results from ATLAS and CMS in this channel can be found in
Ref.~\cite{Aad:2014lma} and Ref.~\cite{Khachatryan:2014qaa} respectively.

Events are selected to contain two high transverse momentum photons and both experiments separate the events from the 8 TeV dataset into two subsamples, one sample with at least one identified lepton
and one sample with zero identified leptons.
ATLAS also performs this separation for the 7 TeV dataset, while CMS merges the zero and one lepton events in the 7 TeV dataset
into a single channel.
For the ATLAS analysis, the events with at least one lepton are required to contain at least one $b$-jet, while the events with zero leptons are required to contain
at least $5$~jets, one of which must be required to be identified as a $b$-jet. For CMS, the events with at least one lepton are required to contain at least two jets, one of which must be identified as a $b$-jet and
the events with zero leptons must contain at least $4$~jets, two of which must be identified as a $b$-jet.

\begin{figure}[b]
\begin{center}
\subfigure[]{
\includegraphics[width=0.48\textwidth]{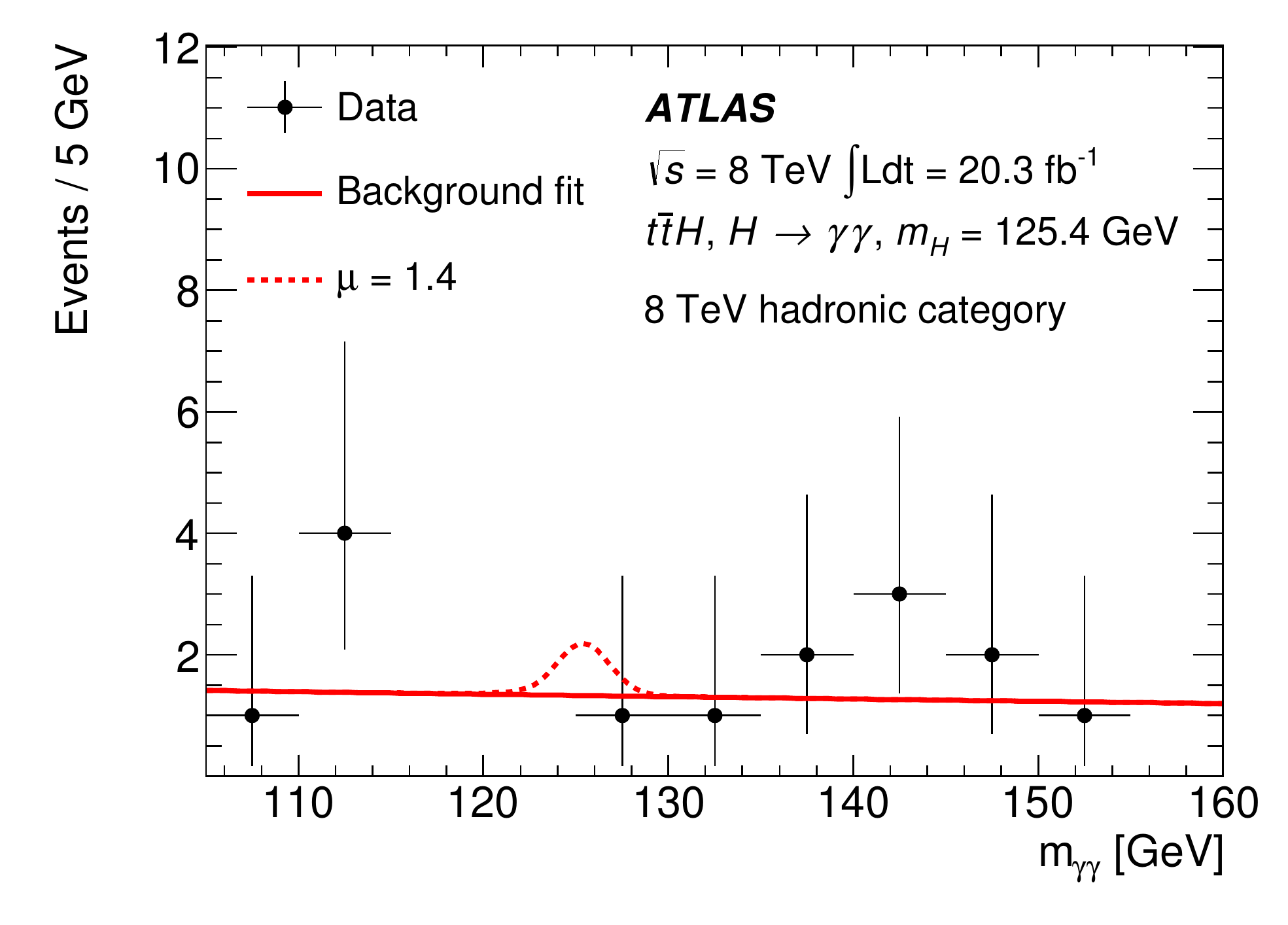}}
\subfigure[]{
\includegraphics[width=0.38\textwidth]{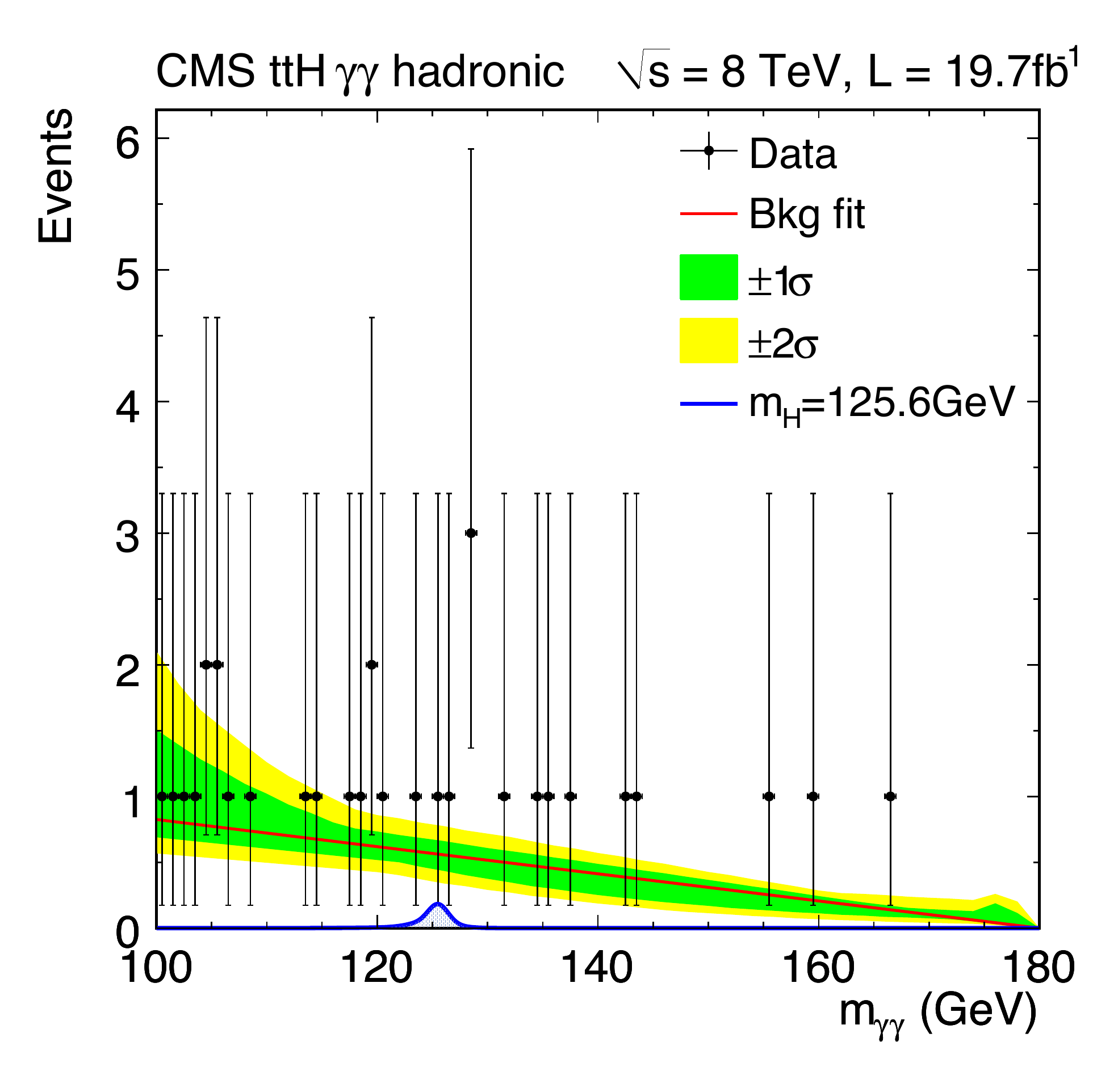}}
\end{center}
\caption{\label{fig-gg-mobs}Observed distribution for the diphoton invariant mass for events in the diphoton search with zero identified leptons in the 8 TeV data for (a) the ATLAS analysis and (b) the CMS analysis.}
\end{figure}

The background is estimated by performing a fit to the observed data. The data observed in the channel with zero identified leptons are shown in
Figure~\ref{fig-gg-mobs}. No significant excess of events is observed in either analysis, and the best fit values for the measured cross-section relative to the SM cross-section are $\mu=1.4^{+2.2}_{-1.4}$~for ATLAS
and $\mu=2.7^{+2.6}_{-1.8}$~for CMS. The (observed) expected $95\%$~CL limits are $6.7\ (4.9)$~times the SM for ATLAS and $7.4\ (4.7)$~times the SM for CMS.

\section{$\ttH$~search in multi-lepton final states}
\label{sec-ml}

Several of the decay modes of the Higgs boson, $H\to W^+W^-$, $H\to ZZ$~and $H\to \tau^+\tau^-$, can produce one or more leptons in the final state.
Once the leptons from the top quark decays are accounted for, this gives rise to final states with multiple leptons and high transverse momentum $b$-jets.
These signatures have a small SM background and so can be used to search for the $\ttH$~process.

CMS has used the following multi-lepton signatures to perform a search for $\ttH$~production~\cite{Khachatryan:2014qaa}: two same-sign charged leptons, three leptons
and four leptons.
The events are required to contain at least two jets and of these jets either one must pass a medium $b$-jet identification requirement or two must pass a loose requirement.
The lepton identification requirements make use of multivariate discriminants; loose requirements are applied to the four lepton channel, while tight requirements
are used in the two and three lepton channels. After applying selection requirements to reject $Z$+jets events, the background is composed of $\ttbar + Z/W^{\pm}$~events,
diboson production, processes with at least one non-prompt lepton and events where the charge of an electron has been mis-measured. The $\ttbar + Z/W^{\pm}$~background
is estimated using MC, but the rate is checked using a control region with three leptons. The diboson background is normalised to the data in a control region with no $b$-jets.
The non-prompt and charge mis-measurement backgrounds are estimated using data driven techniques.

\begin{figure}[b]
\begin{center}
\subfigure[]{
\includegraphics[width=0.23\textwidth]{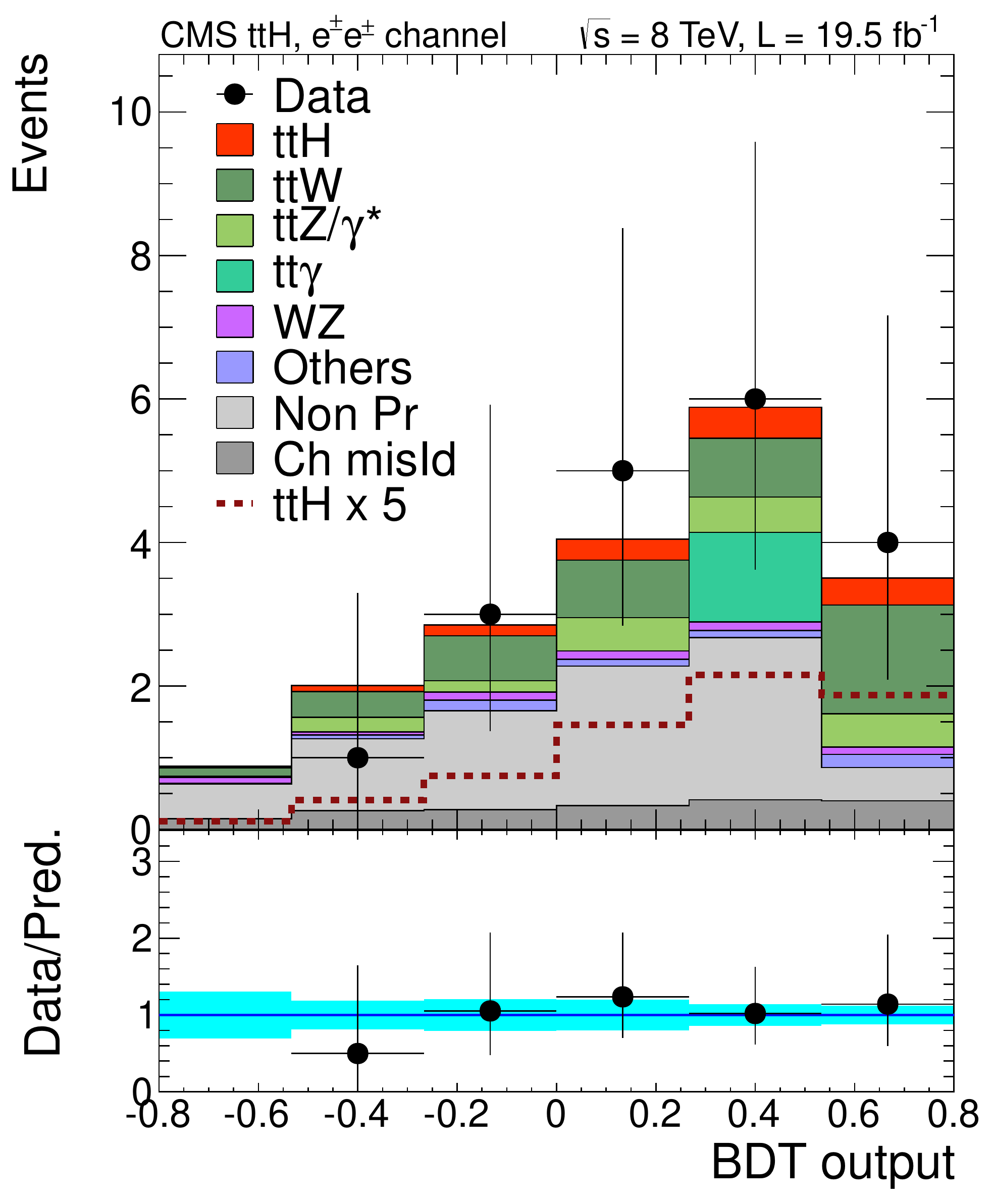}}
\subfigure[]{
\includegraphics[width=0.23\textwidth]{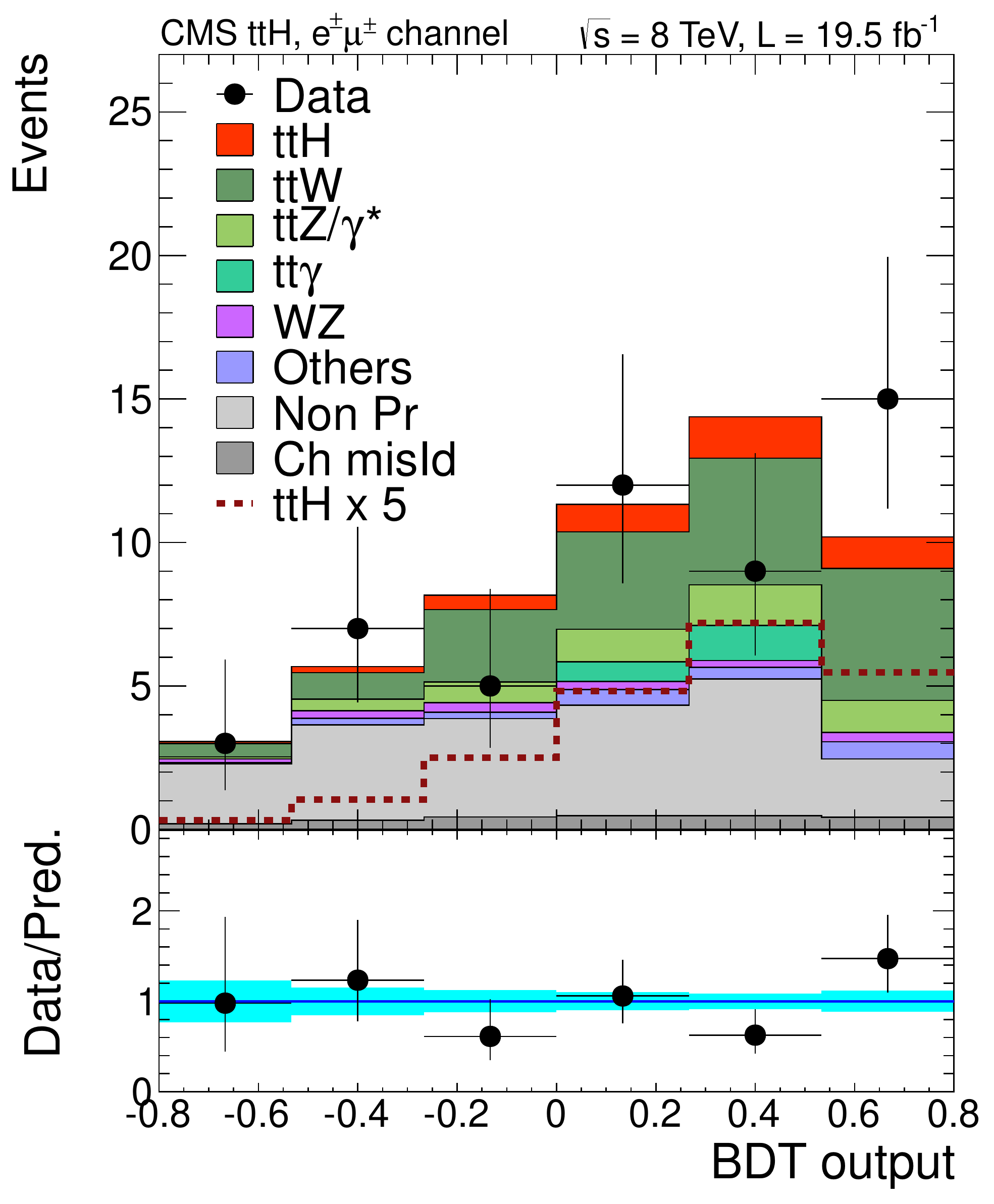}}
\subfigure[]{
\includegraphics[width=0.23\textwidth]{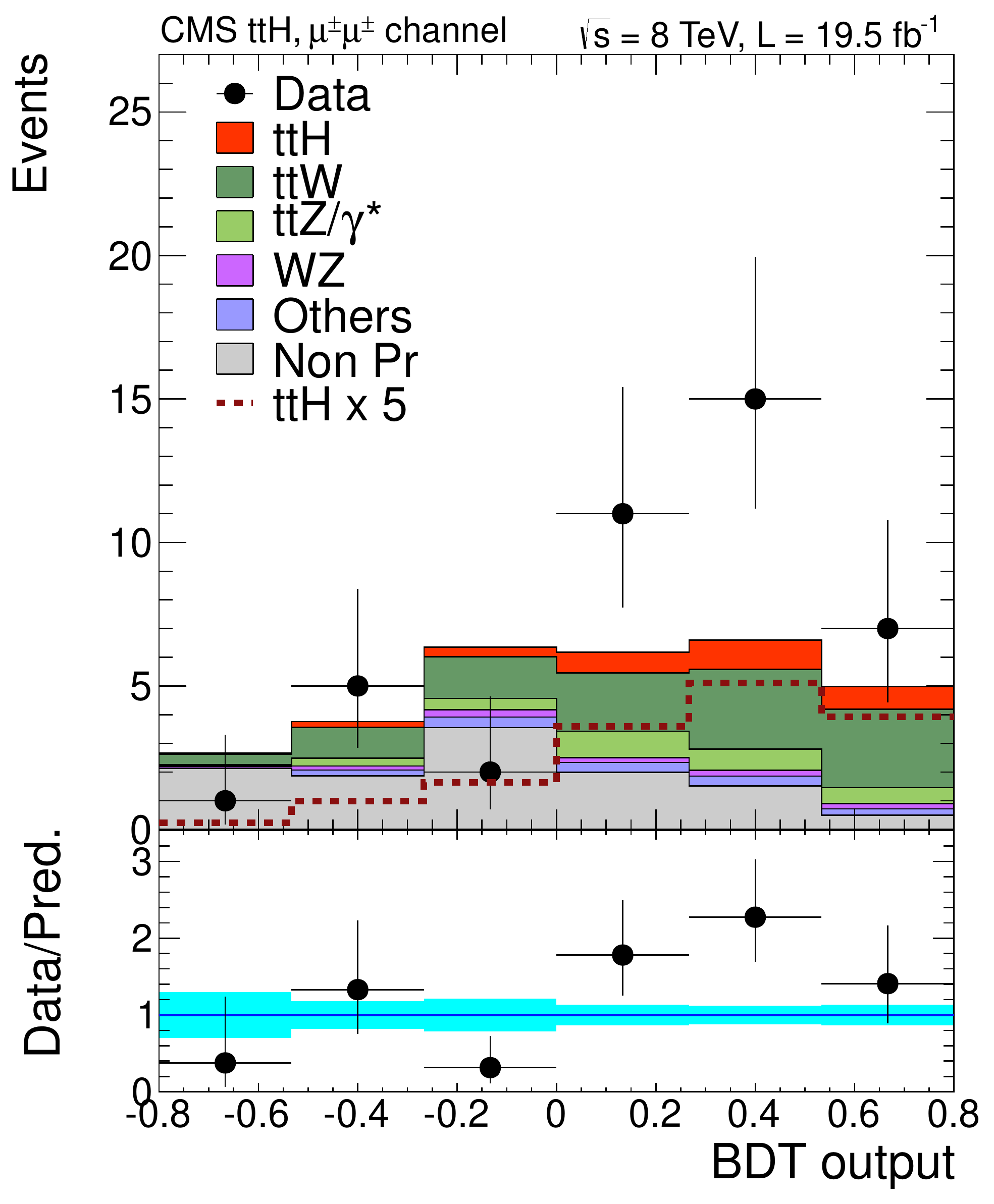}}
\subfigure[]{
\includegraphics[width=0.23\textwidth]{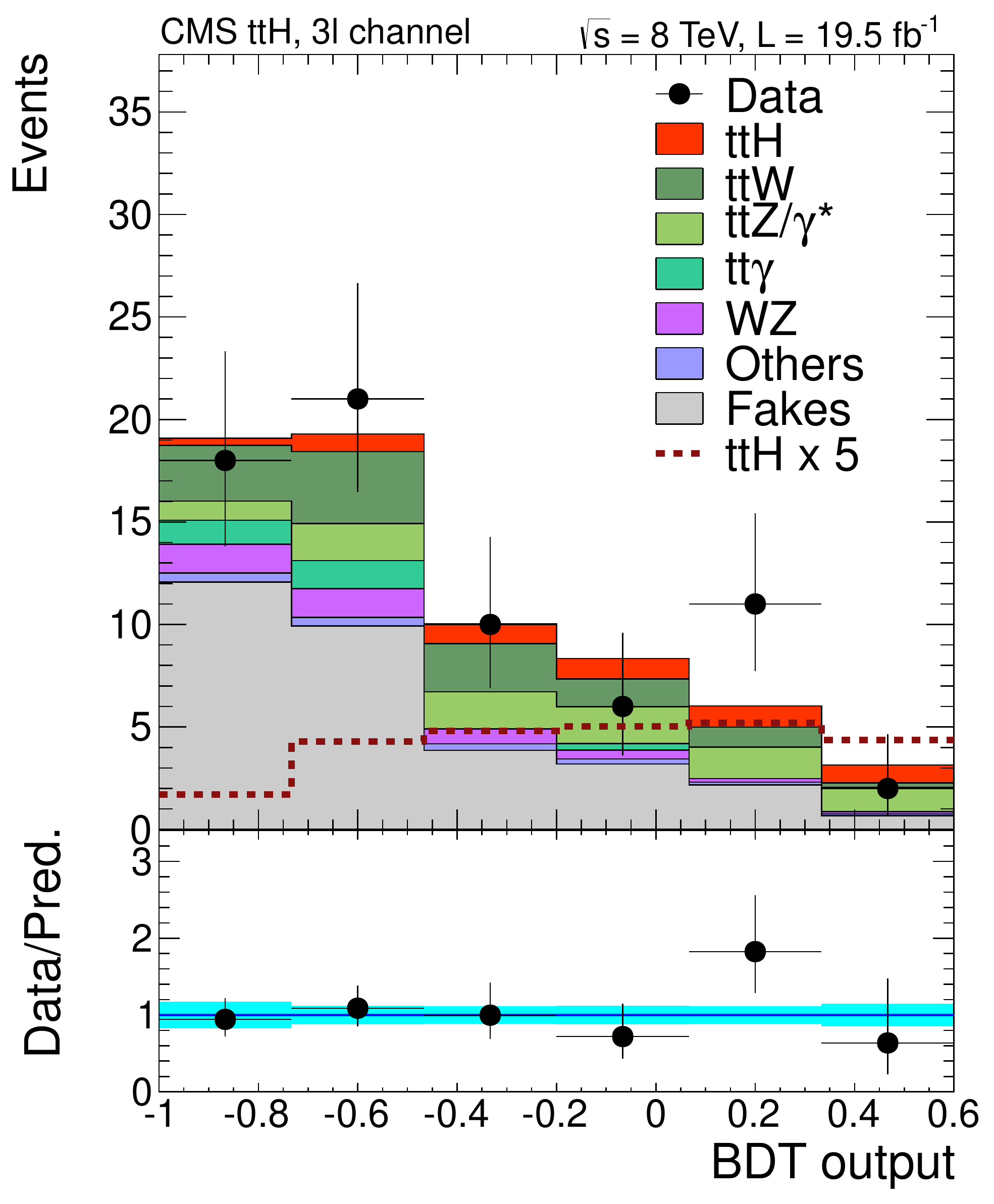}}
\end{center}
\caption{\label{fig-ml-2l}Distribution of the BDT discriminant for (a) $ee$~events, (b) $e\mu$~events, (c) $e\mu$~events
and (d) events with three leptons.}
\end{figure}

To separate the signal from the background, Boosted Decision Trees are trained for the dilepton and trilepton events, using kinematic variables as input. Due to the small rate in the four lepton
channel, the number of jets is used as the discriminating variable. The BDT discriminants are shown for the dilepton and trilepton events in Figure~\ref{fig-ml-2l}.
In most channels, good agreement is seen between the data and the expected backgrounds, however the $\mu\mu$~channel shows an excess of events
that is clustered at high BDT values. The best fit $\mu$~values for the dilepton, trilepton and four lepton channels are $+5.3^{+2.1}_{-1.8}$, $+3.1^{+2.4}_{-2.0}$~and $-4.7^{+5.0}_{-1.3}$. The consistency between the channels is found to be $16\%$.  The large $\mu$~value
in the dilepton channel is driven by the excess of events in the $\mu\mu$~channel seen in Figure~\ref{fig-ml-2l}.

\section{Combination of CMS $\ttH$~searches}

\begin{figure}
\begin{center}
\subfigure[]{
\includegraphics[width=0.43\textwidth]{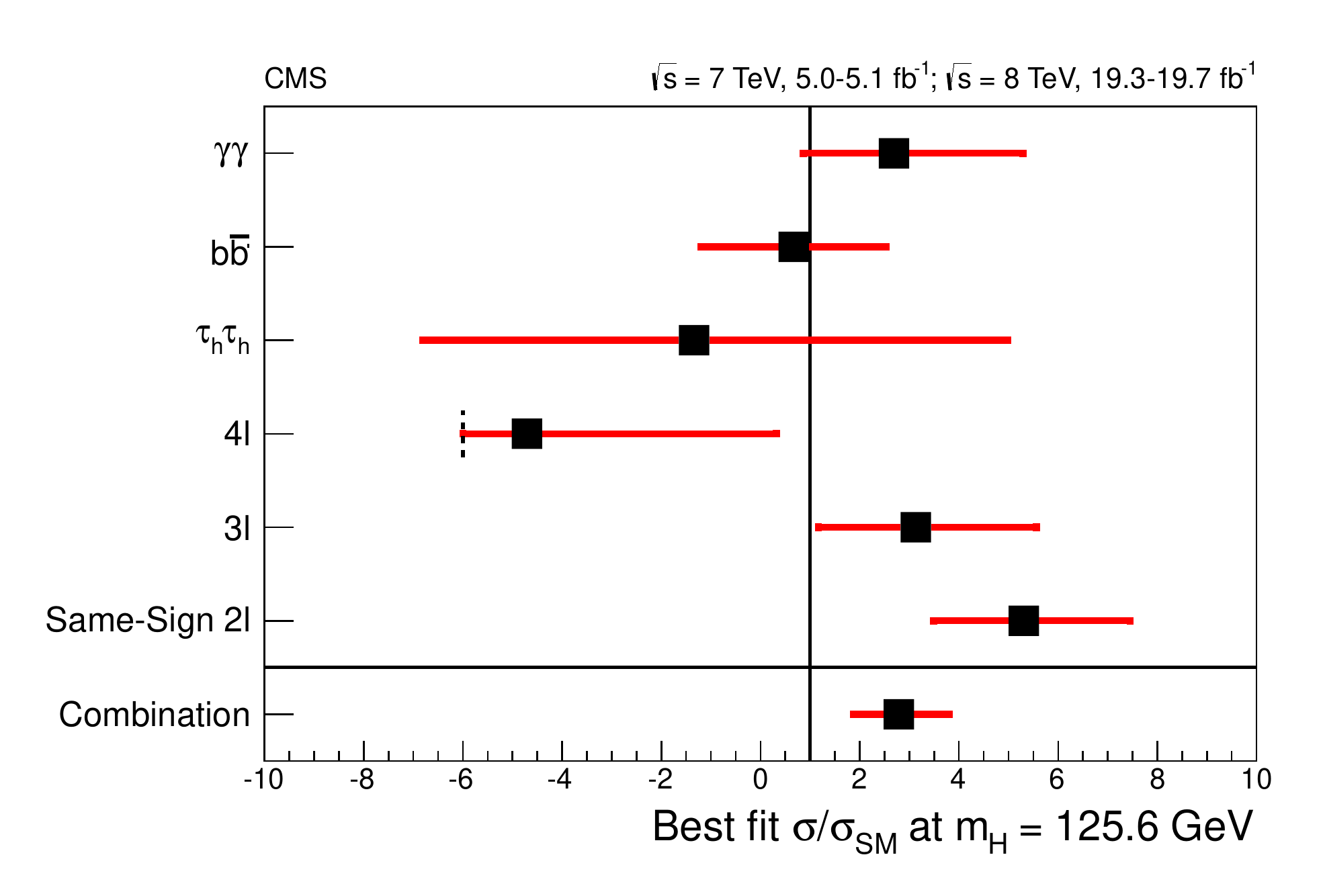}}
\subfigure[]{
\includegraphics[width=0.43\textwidth]{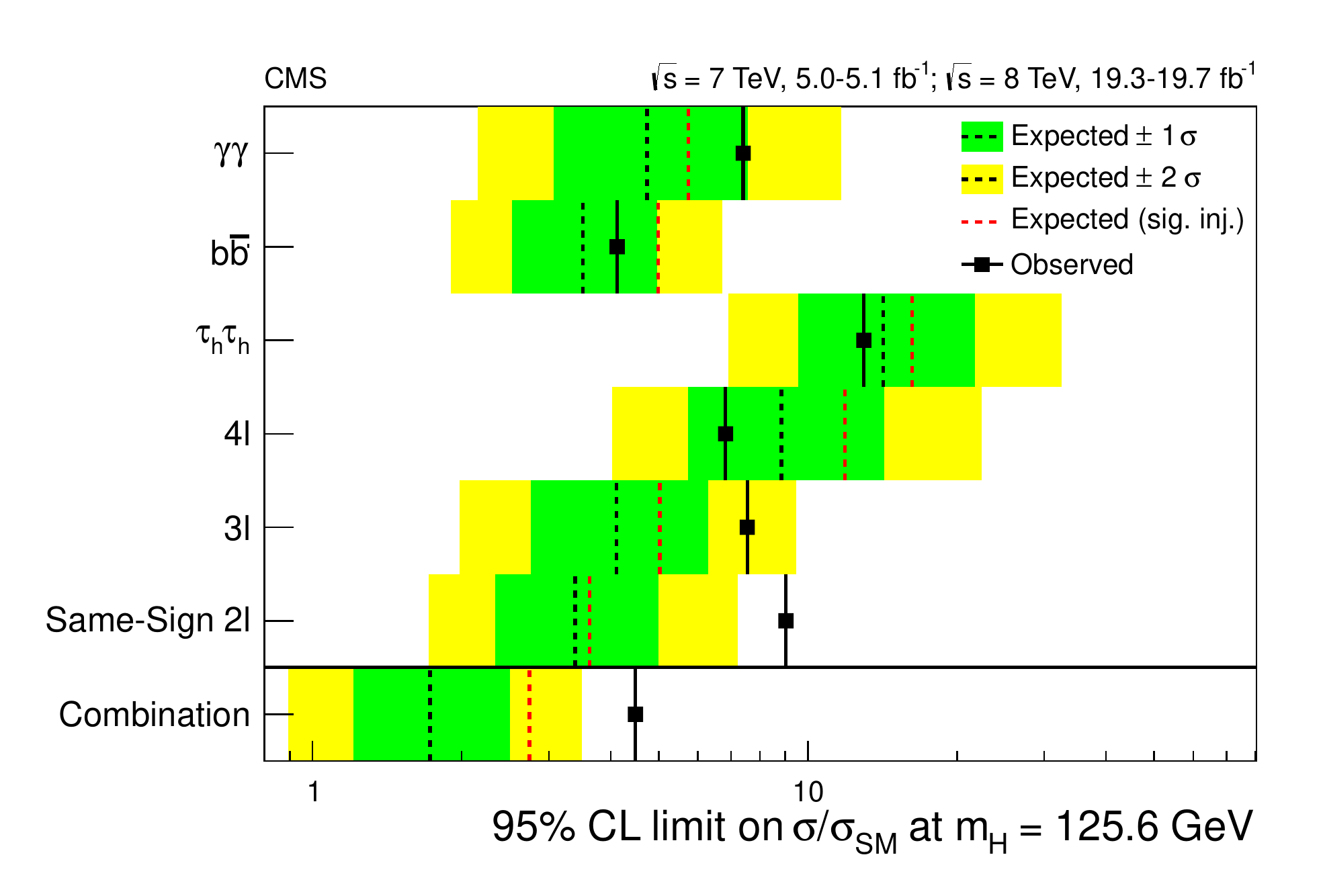}}
\end{center}
\caption{\label{fig-cmscomb} Results of the combined CMS search: (a) the best-fit values of $\mu = \sigma/\sigma_{\mathrm{SM}}$ for each \ttH\ channel at $m_{H}$ = 125.6~GeV are compared to the result from the combination.
(b) The $95\%$~CL upper limits on
      $\mu = \sigma/\sigma_{\mathrm{SM}}$. The black solid and dotted
      lines show the observed and background-only expected limits,
      respectively. The red dotted line shows the median expected
      limit for the SM Higgs boson with $m_{H}$ = 125.6 GeV. The green
      and yellow areas show the $1\sigma$ and $2\sigma$ bands, respectively.}
\end{figure}

CMS has combined the results of the $\ttH$~searches described in Sections~\ref{sec-bb}~to~\ref{sec-ml} in order to maximise the
sensitivity to the $\ttH$~process~\cite{Khachatryan:2014qaa}. In the case of the $H\to \bbbar$~channel, the results from the BDT analysis are used in the combination.
The combination also includes a search for $H\to \tau^+\tau^-$~\cite{Khachatryan:2014qaa}, where the tau leptons decay into hadrons, however this channel
has significantly less sensitivity than the other modes discussed previously.
The combination is performed by making a simultaneous fit to all the channels, correlating the systematic uncertainties where appropriate.
The best fit value for the cross-section relative to the SM cross-section is $\mu=2.8^{+1.0}_{-0.9}$. The combined fit is compared to the fits
in the individual channels in Figure~\ref{fig-cmscomb}. The result represents an excess of  approximately 3.4 standard deviations over
the background-only hypothesis ($\mu=0$)~and 2.1 standard deviations over the expectation for SM $\ttH$~production ($\mu=1$).
The observed and expected $95\%$~CL limits are also shown in Figure~\ref{fig-cmscomb}. This demonstrates the
improvement in the sensitivity of the analysis (the expected limit) obtained by combining the different input channels.

\section{Conclusions}

Searches have been performed for the production of the Higgs boson in association with pairs of top quarks. The searches
have exploited multiple Higgs decay modes, including $H \to \bbbar$, $H \to \gamma\gamma$~$H \to ZZ$,
$H \to W^+W^-$~and $H\to \tau^+\tau^-$. The searches in the $\bbbar$~and $\gamma\gamma$~channels have been performed
by both ATLAS and CMS and no significant excess of events over the background has been observed. The CMS collaboration has
also performed a search using multi-lepton final states (sensitive to $H \to ZZ$, $H \to W^+W^-$~and $H\to \tau^+\tau^-$), where
an excess of events is observed, particularly in events with two same-sign muons. The CMS results in all the channels have been
combined, resulting in a fitted cross-section above the SM expectation.

\end{document}